\documentclass[twocolumn,superscriptaddress,preprintnumbers,amsmath,amssymb,prc]{revtex4-2}

\usepackage{graphicx}
\usepackage{dcolumn}
\usepackage{bm}
\usepackage{ifpdf}
\usepackage{epstopdf}
\usepackage{slashed}
\usepackage{amsfonts}
\usepackage{mathrsfs}
\usepackage{amssymb}
\usepackage{multirow}
\usepackage{CJK}
\usepackage{xcolor}
\usepackage{textcomp}
\usepackage{csquotes}
\usepackage{color}

\begin{document}
\begin{CJK}{UTF8}{ipxm}
\preprint{RIKEN-iTHEMS-Report-23}

\title{Cooper pairing and tripling in one-dimensional spinless fermions with attractive two- and three-body forces}

\author{Yixin Guo (郭一昕)}
\email{guoyixin1997@g.ecc.u-tokyo.ac.jp}
\affiliation{Department of Physics, Graduate School of Science, The University of Tokyo, Tokyo 113-0033, Japan}
\affiliation{RIKEN iTHEMS, Wako 351-0198, Japan}

\author{Hiroyuki Tajima (田島裕之)}
\email{htajima@g.ecc.u-tokyo.ac.jp}
\affiliation{Department of Physics, Graduate School of Science, The University of Tokyo, Tokyo 113-0033, Japan}

\date{\today}

\begin{abstract}
We theoretically investigate
three-body correlations on the top of a Fermi sea in one-dimensional spinless fermions with antisymmetrized two- and three-body attractive interactions.
By investigating the variational problem of three-body states above the Fermi sea, we illuminate the fate of the in-medium three-body cluster states (namely, Cooper triples in the presence of Fermi sea) in the special case with pure attractive three-body interaction as well as in the case with the coexistence of two- and three-body interactions.
Our results testify that the fermion-dimer repulsion is canceled by including the three-body interactions, and stable three-body clusters can be formed.
We further feature a phase diagram consisting of the $p$-wave Cooper pairing and Cooper tripling phases in a plane of $p$-wave two- and three-body coupling strengths.
\end{abstract}

\maketitle

\section{Introduction}\label{sec:I}

A clean and controllable cold-atomic Fermi gas is one of the best candidates to investigate the unconventional states in quantum many-body systems in a systematic way. 
A remarkable feature of this system is the adjustable $s$-wave interaction via the Feshbach resonance~\cite{Chin2010Rev.Mod.Phys.82.1225--1286}.
It is well-known that in a three-dimensional $s$-wave superfluid Fermi gas, the pairing superfluid undergoes a crossover from a Bardeen-Cooper-Schrieffer (BCS) regime with weak-coupling Cooper pairs to a Bose-Einstein condensation (BEC) regime of tightly bound molecules~\cite{doi:10.1146/annurev-conmatphys-031113-133829,Strinati2018Phys.Rep.738.1--76,Ohashi2020Prog.Part.Nucl.Phys.111.103739}. 

The $p$-wave interaction is also tunable near the $p$-wave Feshbach resonance~\cite{Chin2010Rev.Mod.Phys.82.1225--1286}, and a large number of related experiments have been performed towards the realization of $p$-wave Fermi superfluids~\cite{Ticknor2004Phys.Rev.A69.042712,Schunck2005Phys.Rev.A71.045601,Inada2008PhysRevLett.101.100401,Nakasuji2013PhysRevA.88.012710}.
In this regard, an ultracold Fermi gas near the $p$-wave Feshbach resonance may pave a promising way to systematically investigate the role of strong $p$-wave interaction in unconventional superfluids~\cite{Gurarie2007AnnPhys.322.2}.
However, the $p$-wave superfluid has not been achieved experimentally yet due to the difficulty associated with strong atomic losses~\cite{Regal2003Phys.Rev.Lett.90.053201,Zhang2004Phys.Rev.A70.030702,Inada2008PhysRevLett.101.100401,Waseem2017Phys.Rev.A96.062704,Yoshida2018Phys.Rev.Lett.120.133401,Top2021Phys.Rev.A104.043311}.
The $p$-wave Fermi superfluid state is shown to be unstable against three-body clustering in the three-dimensional system even without the Fermi degeneracy~\cite{Levinsen2007PhysRevLett.99.210402}, which leads to the three-body recombination accompanying a strong particle loss.
Indeed, strong three-body losses in three-dimensional Fermi gases near the $p$-wave Feshbach resonance have been observed in experiments~\cite{Waseem2018PhysRevA.98.020702,Yoshida2018Phys.Rev.Lett.120.133401}.
In contrast, the suppression of the three-body loss in the one-dimensional $p$-wave system has been predicted theoretically~\cite{Zhou2017PhysRevA.96.030701}, and a similar result has also been found in our previous work~\cite{Guo2022Phys.Rev.A106.043310}.
In this regard, the systems under the low-dimensional confinement~\cite{Chang2020Phys.Rev.Lett.125.263402} and the lattice geometry~\cite{Venu2023unitary} have been studied experimentally to suppress the atomic loss.
On the other hand, such a suppression of the atomic loss in the one-dimensional system is still under experimental investigation~\cite{Chang2020Phys.Rev.Lett.125.263402,marcum2020suppression,Jackson2022emergent}. 
Indeed, the quasi-one-dimensional geometry may induce the $s$-wave interaction between identical fermions using the orbital degrees of freedom~\cite{Jackson2022emergent}.  Moreover, the three-body clustering with coexistence of $s$- and $p$-wave interactions in one dimension has also been found theoretically~\cite{Guo2023Phys.Rev.B107.024511}.

The stability against the three-body clustering is deeply related to the properties of the interactions and quantum statistics.
While the $s$-wave superconductivity and superfluidity involve the formation of spin-singlet Cooper pairs consisting of two fermions with antiparallel spins due to the fermionic antisymmetrization~\cite{Bardeen1957Phys.Rev.108.1175--1204}, the $p$-wave counterpart can induce the Cooper pairs consisting of two identical fermions in spite of the Pauli exclusion principle. In such a case, one cannot exclude the possibilities of more-than-two-body clustering correlations such as three- and four-body clusters, in contrast with a spin-$1/2$ Fermi gas with strong $s$-wave attractive interaction where the BCS-BEC crossover is realized without any larger clusters due to the Pauli exclusion principle.
To investigate larger clusters, the generalized Cooper problem has been further applied to in-medium cluster states such as Cooper triples~\cite{Niemann2012Phys.Rev.A86.013628,Kirk2017Phys.Rev.A96.053614,Akagami2021Phys.Rev.A104.L041302,Tajima2021Phys.Rev.A104.053328,Tajima2022Phys.Rev.Research4.L012021,Guo2022Phys.Rev.A106.043310,Guo2023Phys.Rev.B107.024511} and even Cooper quartets~\cite{Roepke1998Phys.Rev.Lett.80.3177--3180,Sandulescu2012Phys.Rev.C85.061303,Baran2020Phys.Lett.B805.135462,Guo2022Phys.Rev.C105.024317,Guo2022Phys.Rev.Research4.023152}, which can be regarded as three- and four-body counterparts of a Cooper pair.

Moreover, the three-body force gives a significant influence on the properties of one-dimensional systems.
An important example is the emergence of a quantum anomaly in one-dimensional fermions with the three-body interaction~\cite{Drut2018PhysRevLett.120.243002}, where even an infinitesimally small three-body attraction can cause the three-body bound state.
In such a case, the three-body clusters may survive even in the high-density regime where fermions are quantum degenerate~\cite{McKenney2020Phys.Rev.A102.023313,Tajima2022Phys.Rev.Research4.L012021}. 
Another example is the broken integrability~\cite{PhysRevLett.100.210403}, which is recently discussed in terms of the transport coefficient~\cite{PhysRevE.106.064104}.
Also, the three-body interaction is needed to maintain the Bose-Fermi mapping in one dimension~\cite{Sekino2021Phys.Rev.A103.043307}. 
For the cases of one-dimensional bosonic systems, the three-body interactions lead to the formations of quantum droplet states and excited few-body bound states~\cite{Sekino2018Phys.Rev.A97.011602,Nishida2018Phys.Rev.A97.061603,Pricoupenko2018Phys.Rev.A97.061604,Guijarro2018Phys.Rev.A97.061605}. 
Based on the Bose-Fermi mapping~\cite{Cheon1999Phys.Rev.Lett.82.2536--2539,Valiente2020Phys.Rev.A102.053304,Valiente2021PhysRevA.103.L021302,Sekino2021Phys.Rev.A103.043307}, the counterpart of such unconventional states in bosonic systems are also expected to exist in one-dimensional fermions.
In bosonic systems, the controllable three-body force has been proposed theoretically~\cite{PhysRevLett.112.103201} and demonstrated experimentally~\cite{PhysRevLett.128.083401}. 

While the present authors showed the absence of both in-medium (in the presence of Fermi sea) and in-vacuum three-body bound states in the one-dimensional system with only two-body $p$-wave interaction due to the in-medium fermion-dimer repulsion~\cite{Guo2022Phys.Rev.A106.043310}, if some additional three-body forces exist, the in-vacuum three-body bound state can be induced as found in Ref.~\cite{Sekino2021Phys.Rev.A103.043307}.
Indeed, by including the dimensionless three-body coupling in the atom-dimer scattering, we found the solution of the binding energy for the in-medium three-body bound state in the previous paper~\cite{Guo2022Phys.Rev.A106.043310}.

In this way, it is expected that the three-body ground state can be found by further introducing the three-body interactions.
However, the detailed investigation of the three-body force is still lacking because a specific form of the three-body interaction was phenomenologically introduced in Refs.~\cite{Sekino2021Phys.Rev.A103.043307,Guo2022Phys.Rev.A106.043310}.
In this regard, we start with an antisymmetrized three-body interaction with minimal momentum dependence which can be a leading-order contribution at the low-energy limit, and investigate the three-body clustering in one-dimensional spinless fermions with the coexistence of two- and three-body interactions.
Our results can be further testified in future cold-atomic experiments via three-body loss measurements.
Such an antisymmetrized three-body interaction might be achieved through the quasi-one-dimensionality~\cite{PhysRevE.106.064104} or Rabi coupling~\cite{PhysRevLett.128.083401}. 
In addition, it might also be possible via the medium-induced interaction as we proposed recently by preparing an additional medium~\cite{Guo2023medium}.
While the above approaches were performed in the bosonic systems, they are also expected to be realized in the fermionic ones.
The systematic studies of the effects of two- and three-body forces in fermionic systems would also be useful for understanding the role of three-body forces in nuclear systems~\cite{RevModPhys.85.197},
which have been examined in recent experiments~\cite{PhysRevC.103.044001}.

This paper is organized as follows:
In Sec.~\ref{sec:II}, we first introduce the Hamiltonian for the one-dimensional spinless fermions with the coexistence of two- and three-body forces.
After that, we calculate the expectation value of the energy and derive the corresponding variational equation.
The results and discussion will be given in Sec.~\ref{sec:III}.
In detail, we first investigate the in-medium three-body problem in one-dimensional spinless $p$-wave fermions with pure three-body interaction in Sec.~\ref{sec:IIIA}.
As a step further, by also including the two-body interaction, we study the in-medium three-body clustering in the general case with the coexistence of two- and three-body interactions in Sec.~\ref{sec:IIIB}.
Finally, we summarize this paper in Sec.~\ref{sec:IV}.
In the following, we take $\hbar=c=k_{\rm B}=1$.
The system size is taken to be unity.

\section{Theoretical framework}\label{sec:II}

We consider one-dimensional spinless fermions with two- and three-body interactions described by the Hamiltonian
\begin{align}\label{hamiltonian}
H=K+V_2+V_3,
\end{align}
where the kinetic term $K$ and two-body interaction $V_2$ are given as
\begin{align}
K=\,&\sum_{k}\xi_{k}c_{k}^\dag c_{k},\\
    V_2=\,&\frac{U_2}{2}\sum_{k_1,k_2,k_1',k_2'}
    \left(\frac{k_1-k_2}{2}\right)
    \left(\frac{k_1'-k_2'}{2}\right)\nonumber\\
    &\times B_{k_1,k_2}^\dag
    B_{k_2',k_1'}
    \delta_{k_1+k_2,k_1'+k_2'},
\end{align}
respectively. 
Here, $\xi_{k}=k^2/(2m)-\mu$ is the single-particle energy with momentum $k$, atomic mass $m$, and chemical potential $\mu$.
The two-body interaction adopted here corresponds to the short-range $p$-wave type with a coupling constant $U_2$, which is related to the zero-range limit of the two-channel model for the Feshbach resonance~\cite{Cui2016PhysRevA.94.043636,Tajima2021PhysRevA.104.023319}.
The relation between $U_2$ and the $p$-wave scattering length $a$ is obtained from the two-body $T$ matrix as~\cite{Cui2016PhysRevA.94.043636,Valiente2020Phys.Rev.A102.053304}
\begin{align}
    \frac{1}{U_2}-\sum_{p}\frac{mp^2}{k^2+i\delta-p^2}
    =\frac{m}{2}\left(\frac{1}{a}-\frac{1}{2}r_{\rm eff}k^2+ik\right),
\end{align}
where $r_{\rm eff}$ is the effective range and $\delta$ is an infinitesimally small number.
$r_{\rm eff}$ is associated with the momentum cutoff $\Lambda=\frac{4}{\pi r_{\rm eff}}$.
Taking the consideration of the antisymmetry and parity for the form factor, we introduce an antisymmetrized attractive three-body interaction as
\begin{align}
V_3=\,&\sum_{k_1,k_2,k_3}\sum_{k_1',k_2',k_3'}U_3\nonumber\\
    &\times\left(\frac{k_1-k_2}{2}\right)
    \left(\frac{k_2-k_3}{2}\right)
        \left(\frac{k_3-k_1}{2}\right)
    \nonumber\\
    &
    \times
      \left(\frac{k_1'-k_2'}{2}\right)
    \left(\frac{k_2'-k_3'}{2}\right)    
    \left(\frac{k_3'-k_1'}{2}\right)\nonumber\\
    &\times    
    c_{k_1}^\dag  c_{k_2}^\dag c_{k_3}^\dag c_{k_3'}  c_{k_2'}c_{k_1'}\delta_{k_1+k_2+k_3,k_1'+k_2'+k_3'},
\end{align}
where $U_3$ is the coupling constant of the three-body interaction.
Moreover,
by introducing the pair operator
\begin{align}
    B_{k_1,k_2}^\dag = c_{k_1}^\dag c_{k_2}^\dag, \quad B_{k_1,k_2}= c_{k_2}c_{k_1},
\end{align}
and the trimer operator
\begin{align}
    F_{k_1,k_2,k_3}^\dag = c_{k_1}^\dag c_{k_2}^\dag c_{k_3}^\dag, \quad F_{k_1,k_2,k_3} = c_{k_3}c_{k_2}c_{k_1},
\end{align}
one can rewrite $V_3$ as
\begin{align}
    V_3=\,&\sum_{k_1,k_2,k_3}\sum_{k_1',k_2',k_3'}U_3\nonumber\\
    &\times\left(\frac{k_1-k_2}{2}\right)
    \left(\frac{k_2-k_3}{2}\right)
        \left(\frac{k_3-k_1}{2}\right)
    \nonumber\\
   &\times 
   \left(\frac{k_1'-k_2'}{2}\right)
   \left(\frac{k_2'-k_3'}{2}\right)
    \left(\frac{k_3'-k_1'}{2}\right)\nonumber\\
 &\times   B_{k_1,k_2}^\dag c_{k_3}^\dag c_{k_3'} B_{k_1',k_2'}\delta_{k_1+k_2+k_3,k_1'+k_2'+k_3'}  \nonumber\\
    =\,& \sum_{k_1,k_2,k_3}\sum_{k_1',k_2',k_3'}U_3\nonumber\\
   &\times \left(\frac{k_1-k_2}{2}\right)
     \left(\frac{k_2-k_3}{2}\right)
          \left(\frac{k_3-k_1}{2}\right)
     \nonumber\\
   &\times 
   \left(\frac{k_1'-k_2'}{2}\right)
   \left(\frac{k_2'-k_3'}{2}\right)
    \left(\frac{k_3'-k_1'}{2}\right)\nonumber\\
&\times    F_{k_1,k_2,k_3}^\dag F_{k_1',k_2',k_3'}\delta_{k_1+k_2+k_3,k_1'+k_2'+k_3'}.
    \end{align}
Such a kind of three-body interaction would be the leading order of the antisymmetrized attractive ones at low energy in the sense of the derivative expansion.

We note that the two- and three-body interactions considered in the Hamiltonian~\eqref{hamiltonian} can also be related to the scattering hypervolume $D_F$~\cite{Tan2008Phys.Rev.A78.013636}, which is a three-body analog of the two-body scattering length.
In general, $D_{F}$ can be extracted by solving the three-body Schr\"odinger equation numerically at zero energy and matching the resultant wave function with the asymptotic expansions of wave function~\cite{dftan2023}.
In this regard, since the calculation of $D_F$ for given $U_2$ and $U_3$ is not so straightforward and moreover $D_F$ has not been experimentally measured yet in contrast to the two-body scattering length,
we measure $U_3$ by using the critical coupling strength $U_{\rm c}$ where the three-body bound state appears only with the three-body attraction in vacuum.
Accordingly, $U_{\rm c}$ may be regarded as the resonant coupling where $D_{F}$ diverges~\cite{dftan2023}.

In a way similar to the previous works~\cite{Kirk2017Phys.Rev.A96.053614,Tajima2021Phys.Rev.A104.053328, Guo2022Phys.Rev.A106.043310}, the trial wave function for the three-body cluster on the top of the Fermi sea is given by
\begin{align}\label{twf}
    |\Psi_{3}\rangle=\sum_{p_1,p_2,p_3}'\delta_{p_1+p_2,-p_3}\Omega_{p_1,p_2}F_{p_1,p_2,p_3}^\dag |{\rm FS}\rangle,
\end{align}
where $\Omega_{p_1,p_2}$ is the variational parameter and the three-body state with zero center-of-mass momentum ($p_1+p_2+p_3=0$) is considered.
Hereafter, we introduce the momentum summation restricted by the Pauli blocking as
\begin{align}
    &\sum_{k_1,k_2,\cdots}'\mathcal{F}(k_1,k_2,\cdots)\nonumber\\
    =\,&\sum_{k_1,k_2,\cdots}\theta(|k_1|-k_{\rm F})\theta(|k_2|-k_{\rm F})\cdots  F(k_1,k_2,\cdots),
\end{align}
for an arbitrary function $\mathcal{F}(k_1,k_2,\cdots)$, where
$k_{\rm F}=\sqrt{2mE_{\rm F}}$ is the Fermi momentum.

\begin{widetext}
From the variational principle with the in-medium three-body energy $E_3$, we obtain
\begin{align}
\frac{\delta\langle\Psi_3|(H-E_3)|\Psi_3\rangle}{\delta\Omega_{p_1,p_2}^*}=\frac{\delta\langle\Psi_3|(K+V_2+V_3-E_3)|\Psi_3\rangle}{\delta\Omega_{p_1,p_2}^*}=0.
\end{align}
Detailed results for the expectation values of each term in the Hamiltonian are given in Appendix~\ref{appendixA}.
The resulting variational equation reads
\begin{align}\label{veq}
    &2(\xi_{p_1}+\xi_{p_2}+\xi_{p_3}-E_3)\left[\Omega_{p_1,p_2}+\Omega_{p_2,p_3}+\Omega_{p_3,p_1}\right]\cr
    =\,&-\frac{U_2}{2}\sum_{q}'\left[(p_1-p_2)(2q+p_3)(2\Omega_{p_3,q}+\Omega_{q,-q-p_3})+(p_2-p_3)(2q+p_1)(2\Omega_{p_1,q}+\Omega_{q,-q-p_1})
    \right.\cr
    &+\left.(p_3-p_1)(2q+p_2)(2\Omega_{p_2,q}+\Omega_{q,-q-p_2})\right]\nonumber\\
    &-\frac{9U_3}{16}
\sum_{p'_1,p'_2}'
\left({p_1-p_2}\right)
\left({p_1'-p_2'}\right)
\left({p_1+2p_2}\right)
\left({p_1'+2p_2'}\right)
\left({2p_1+p_2}\right)
\left({2p_1'+p_2'}\right)
\Omega_{p'_1,p'_2}.
\end{align}
\end{widetext}

\section{Results and Discussion}\label{sec:III}

\subsection{Pure three-body interaction (without two-body interaction)}\label{sec:IIIA}

First, let us consider the case without two-body interaction ($U_2=0$).
In this case, the variational equation~\eqref{veq} can be recast into
\begin{align}
        &2(\xi_{p_1}+\xi_{p_2}+\xi_{p_3}-E_3)\left[\Omega_{p_1,p_2}+\Omega_{p_2,p_3}+\Omega_{p_3,p_1}\right]\nonumber\\
    &\ \ = 
    \frac{9U_3}{16}
    \mathcal{X}_{p_1,p_2,p_3}
    \mathcal{C},
\end{align}
where for convenience we introduced 
\begin{align}
\mathcal{X}_{p_1,p_2,p_3}=(p_1-p_2)(p_2-p_3)(p_3-p_1),
\end{align}
and
\begin{align}\label{eqC}
    \mathcal{C}
&=-\sum_{p'_1,p'_2,p_3'}'\delta_{p_1'+p_2'+p_3',0}
\mathcal{X}_{p'_1,p_2',p_3'}
\Omega_{p'_1,p'_2}.
\end{align}
Consequently, we can obtain the equation regarding the amplitude of the trial wave function $\Omega_{p,k}$ as
\begin{align}
        &
        \mathcal{X}_{p_1,p_2,p_3}
        [\Omega_{p_1,p_2}+\Omega_{p_2,p_3}+\Omega_{p_3,p_1}]\cr
    & \ \ \ =
    \frac{9U_3}{16}
    \frac{
    \mathcal{X}_{p_1,p_2,p_3}^2
    \mathcal{C}}{2(\xi_{p_1}+\xi_{p_2}+\xi_{p_3}-E_3)}.
\end{align}
Finally, by further taking the momenta summations with $-\delta_{p_1+p_2+p_3,0}$, the three-body equation with only the three-body interaction reads
\begin{align}\label{pure3b}
    1=-\frac{3U_3}{16}\sum_{p_1,p_2}'
    \frac{
    \mathcal{X}_{p_1,p_2,p_3}^2
    }{2(\xi_{p_1}+\xi_{p_2}+\xi_{p_3}-E_3)},
\end{align}
where the constraint $|p_3|=|-p_1-p_2|\geq k_{\rm F}$ is used in Eq.~\eqref{pure3b}.

\begin{figure}
  \includegraphics[width=0.45\textwidth]{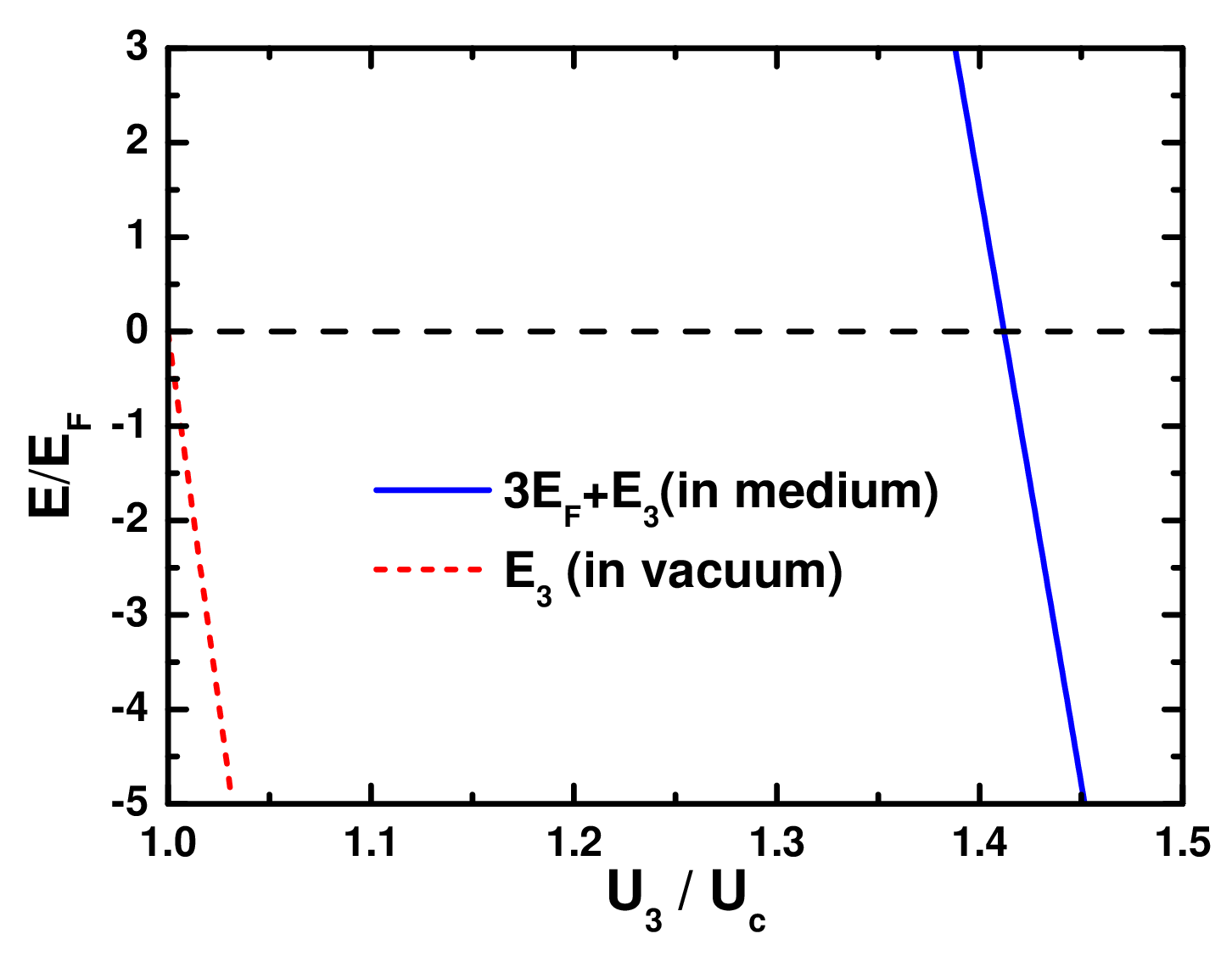}
  \caption{ 
  The relation between three-body state energy 
  $E_3$ and three-body interaction strength ${U}_3$. 
  The critical coupling strength $U_c$ is a coupling strength where the three-body bound state appears in vacuum.
  In this figure, $\Lambda$ is taken as $10k_{\rm F}$.
  }\label{fig:1}
\end{figure}

The solution of the in-medium three-body energy in the case with the pure three-body interaction~\eqref{pure3b} is shown in the solid line in Fig.~\ref{fig:1} with a shift of $3E_{\rm F}$, where  the ultraviolet momentum cutoff $\Lambda$ is taken as $10k_{\rm F}$.
For comparison, the solution of the in-vacuum three-body energy is also plotted in the dashed line.
The three-body coupling constant $U_3$ is normalized by the in-vacuum critical coupling $U_{\rm c}$, where the three-body bound state starts to appear in vacuum.
The results indicate that the in-medium three-body bound state assisted by the Fermi-surface effect does not exist in the absence of the in-vacuum counterpart, which is different from the three-component Fermi gas with two-body $s$-wave interactions in three dimensions~\cite{Tajima2021Phys.Rev.A104.053328}.
However, there is still a regime (around $1.38<{U}_{3}/U_c<1.41$) with positive $E_3+3E_{\rm F}$ in medium, which corresponds to the so-called squeezed Cooper triple regime in the three-dimensional $s$-wave case~\cite{Tajima2021Phys.Rev.A104.053328}.
In analogy with the pairing state in the BCS-BEC crossover~\cite{Strinati2018Phys.Rep.738.1--76,Ohashi2020Prog.Part.Nucl.Phys.111.103739}, actually there is no noticeable structural change of Cooper triples.
In this sense, the regime with positive $E_3+3E_{\rm F}$ in medium (around $1.38<{U}_{3}/U_c<1.41$) in Fig.~\ref{fig:1} can be still regarded as the Cooper triple phase but may have a relatively smaller cluster size due to a strong attraction as to ensure a three-body cluster would be bound in vacuum.
We note that the in-medium three-body bound states do not always require a stronger three-body coupling than the case in vacuum, and one can find the loosely-bound Cooper triples due to the interplay between two- and three-body interactions, which is shown in the latter section.

Finally, the in-medium three-body binding energy $E_3+3E_{\rm F}$ turns to negative when ${U}_{3}/U_c$ is around $1.41$.
Such a regime corresponds to the formation of three-body bound states which are dominated by the strong three-body interaction.
The coupling strength ${U}_{3}/U_c\simeq 1.41$, where $E_3+3E_{\rm F}=0$, is analogous to the region where the chemical potential changes the sign in the BCS-BEC crossover~\cite{Strinati2018Phys.Rep.738.1--76,Ohashi2020Prog.Part.Nucl.Phys.111.103739}.
Although in this work we consider a three-body system on top of the Fermi sea, our results can describe the qualitative features of both the weak- and strong-coupling limits appropriately by the variational equation in a unified manner~\cite{Tajima2021Phys.Rev.A104.053328}. 

\subsection{Coexistence of two- and three-body interactions}\label{sec:IIIB}

In this subsection, we further investigate the general case with the presence of both two- and three-body interactions.
To simplify the expressions, besides $\mathcal{C}$ as given in Eq.~\eqref{eqC}, here we also introduce 
\begin{align}
    \mathcal{A}(p_1,p_2)=\sum_{q}'(p_1-p_2)(2q+p_3)(2\Omega_{p_3,q}+\Omega_{q,-q-p_3}),
\end{align}
\begin{align}
    \mathcal{B}(p_2)&=\sum'_{p_1,p_3}
    (\Omega_{p_1,p_2}+\Omega_{p_3,p_1}+\Omega_{p_2,p_3})\cr
    &\ \times
    (p_3-p_1)
    \delta_{p_1+p_2+p_3,0},
\end{align}
where
\begin{align}
    \mathcal{A}(p_1,p_2)=(p_1-p_2)\mathcal{B}(p_3).
\end{align}
\begin{widetext}
With the help of above notations,  we obtain the closed equation for $\mathcal{B}(p)$ and $E_3$ as
\begin{align}\label{closedeq}
    &\mathcal{B}(p_2)
    \left[\frac{1}{U_2}+\sum'_{p_1}\frac{(p_1+p_2/2)^2}{\xi_{p_1}+\xi_{p_2}+\xi_{p_3}-E_3}\right]
    =\sum'_{p_1}\frac{(p_1+2p_2)(p_1+p_2/2)\mathcal{B}(p_1)
    }{\xi_{p_1}+\xi_{p_2}+\xi_{p_3}-E_3}\cr
& \ \ +\frac{\frac{9U_3}{16}\sum'_{p_1}\frac{
\mathcal{X}_{p_1,p_2,p_3}
\left({p_3-p_1}\right)}{2(\xi_{p_1}+\xi_{p_2}+\xi_{p_3}-E_3)}}{1+\frac{3U_3}{16}\sum'_{p_1,p_2}
    \frac{\mathcal{X}_{p_1,p_2,p_3}^2
    }{2(\xi_{p_1}+\xi_{p_2}+\xi_{p_3}-E_3)}}
    \sum'_{p_1,p_2}\frac{(p_2+p_1/2){\mathcal{X}_{p_1,p_2,p_3}}\mathcal{B}(p_1)}{2(\xi_{p_1}+\xi_{p_2}+\xi_{p_3}-E_3)},
\end{align}
from the full variational equation~\eqref{veq}.
In Eq.~\eqref{closedeq}, we take $p_3=-p_1-p_2$ because of the momentum conservation at the center-of-mass frame of the three-body system as in Eq.~\eqref{pure3b}.
The detailed derivations for Eq.~\eqref{closedeq} can be found in Appendix~\ref{appendixB}.
\end{widetext}
For convenience, we rewrite Eq.~\eqref{closedeq} as
\begin{align}\label{3beq}
    \mathcal{B}(p_2)\left[\frac{1}{U_2}+I_2(p_2,E_3)\right]
    =I_3(p_2,E_3)+\frac{I_4(E_3)I_5(p_2,E_3)}{1+I_6(E_3)},
\end{align}
where the integrals read
\begin{align}
    I_2(p_2,E_3)=
    \sum'_{p_1}
    \frac{(p_1+p_2/2)^2}{\xi_{p_1}+\xi_{p_2}+\xi_{p_3}-E_3},
\end{align}
\begin{align}
    &I_3(p_2,E_3)=
    \sum'_{p_1}\frac{(p_1+p_2/2)(p_1+2p_2)
    \mathcal{B}(p_1)}{\xi_{p_1}+\xi_{p_2}+\xi_{p_3}-E_3},
\end{align}
\begin{align}
I_4(E_3)=\sum'_{p_1,p_2}\frac{(p_2+p_1/2)\mathcal{X}_{p_1,p_2,p_3}\mathcal{B}(p_1)}{2(\xi_{p_1}+\xi_{p_2}+\xi_{p_3}-E_3)},
\end{align}
\begin{align}
    I_5(p_2,E_3)=\frac{9U_3}{16}\sum'_{p_1}\frac{
\mathcal{X}_{p_1,p_2,p_3}
\left({p_3-p_1}\right)
}{2(\xi_{p_1}+\xi_{p_2}+\xi_{p_3}-E_3)},
\end{align}
and
\begin{align}
    I_6(E_3)=\frac{3U_3}{16}\sum'_{p_1,p_2}
    \frac{
    \mathcal{X}_{p_1,p_2,p_3}^2
    }{2(\xi_{p_1}+\xi_{p_2}+\xi_{p_3}-E_3)},
\end{align}
respectively, where we apply the constraints $|p_3|\equiv|p_1+p_2|\geq k_{\rm F}$ in each momentum summation.
We note that $1+I_6(E_3)=0$ corresponds to the three-body equation~\eqref{pure3b} for the case with pure three-body interaction discussed in Sec.~\ref{sec:IIIA}.

\begin{widetext}
In addition, the right-hand side of Eq.~\eqref{closedeq} can be recast into
\begin{align}
\label{eq:v3}
    &\sum_{p_1}'\frac{(p_1+2p_2)(p_1+p_2/2)\mathcal{B}(p_1)
    }{\xi_{p_1}+\xi_{p_2}+\xi_{p_3}-E_3}
+\frac{\frac{9U_3}{16}\sum_{p_1}'\frac{
\mathcal{X}_{p_1,p_2,p_3}
\left({p_3-p_1}\right)}{2(\xi_{p_1}+\xi_{p_2}+\xi_{p_3}-E_3)}}{1+\frac{3U_3}{16}\sum_{p_1,p_2}'
    \frac{\mathcal{X}_{p_1,p_2,p_3}^2
    }{2(\xi_{p_1}+\xi_{p_2}+\xi_{p_3}-E_3)}}
{\sum_{p_1,p_2}'\frac{(p_2+p_1/2)\mathcal{X}_{p_1,p_2,p_3}\mathcal{B}(p_1)}{2(\xi_{p_1}+\xi_{p_2}+\xi_{p_3}-E_3)}}\nonumber\\
    =\,&\sum_{p_1}'\left[
    \frac{(p_1+2p_2)(p_1+p_2/2)
    }{\xi_{p_1}+\xi_{p_2}+\xi_{p_3}-E_3}
+\frac{\frac{9U_3}{16}\sum_{p_1}'\frac{
\mathcal{X}_{p_1,p_2,p_3}
\left({p_3-p_1}\right)}{2(\xi_{p_1}+\xi_{p_2}+\xi_{p_3}-E_3)}}{1+\frac{3U_3}{16}\sum_{p_1,p_2}'
    \frac{\mathcal{X}_{p_1,p_2,p_3}^2
    }{2(\xi_{p_1}+\xi_{p_2}+\xi_{p_3}-E_3)}}
    {\sum_{p_2}'\frac{(p_2+p_1/2){\mathcal{X}_{p_1,p_2,p_3}}}{2(\xi_{p_1}+\xi_{p_2}+\xi_{p_3}-E_3)}}\right]\mathcal{B}(p_1)\nonumber\\
    \equiv\,&-\frac{m}{2}\sum_{p_1}'\left[t_{\rm F}(p_1,p_2,E_3)+\mathcal{V}_3(p_1,p_2,E_3)\right]\mathcal{B}(p_1),
\end{align}
where
\begin{align}
    t_{\rm F}(p_1,p_2,E_3)=-\frac{2}{m}\frac{(p_1+2p_2)(p_1+p_2/2)
    }{\xi_{p_1}+\xi_{p_2}+\xi_{p_3}-E_3},
    \end{align}
    corresponds to Eq.~(38) in Ref.~\cite{Guo2022Phys.Rev.A106.043310} and
    \begin{align}\label{eqv3}
    \mathcal{V}_3(p_1,p_2,E_3)\equiv-\frac{2}{m}\frac{\frac{9U_3}{16}\sum_{p_1}'\frac{
\mathcal{X}_{p_1,p_2,p_3}
\left({p_3-p_1}\right)}{2(\xi_{p_1}+\xi_{p_2}+\xi_{p_3}-E_3)}}{1+\frac{3U_3}{16}\sum_{p_1,p_2}'
    \frac{\mathcal{X}_{p_1,p_2,p_3}^2
    }{2(\xi_{p_1}+\xi_{p_2}+\xi_{p_3}-E_3)}}
{    \sum_{p_2}'\frac{(p_2+p_1/2)\mathcal{X}_{p_1,p_2,p_3}}{2(\xi_{p_1}+\xi_{p_2}+\xi_{p_3}-E_3)}}.
\end{align}
From the above discussion, one can find that $\mathcal{V}_3(p_1,p_2,E_3)$ corresponds to the dimensionless constant three-body (atom-dimer) coupling $\mathcal{V}_3=2$ introduced in Refs.~\cite{Sekino2021Phys.Rev.A103.043307,Guo2022Phys.Rev.A106.043310}, while instead $\mathcal{V}_3(p_1,p_2,E_3)$ has the momentum and energy dependence.
\end{widetext}

In the practical calculation, we solve Eq.~\eqref{3beq} with an iteration method numerically evaluating ${I}_2(p_2,E_3)$, ${I}_3(p_2,E_3)$, {${I}_4(E_3)$}, ${I}_5(p_2,E_3)$, and ${I}_6(E_3)$. 
We start the iteration from the initial value $\mathcal{B}({p})=1$ for a given value of $E_3$, since $E_3$ is unchanged by the scale transformation of the solution $\mathcal{B}({p})$ in Eq.~\eqref{3beq}. 
For the convergence of $\mathcal{B}({p})$, we have required
\begin{align}
  \sum_n\frac{\left[\mathcal{B}_{\rm {in }}(n)-\mathcal{B}_{\rm {out }}(n)\right]^2}{\mathcal{B}_{\rm {in }}(n)^2} \leqslant 10^{-8},
\end{align}
where $\mathcal{B}_{\rm {in}}(n)$ [$\mathcal{B}_{\rm {out}}(n)$] is the input (output) during the iteration for Eq.~\eqref{3beq}.
$n$ is the number of the discretized momentum $p=n\Delta p + k_{\rm F}$ used in the Newton-Cotes integration with $\Delta p=(\Lambda-k_{\rm F})/N$. 
We have confirmed that $N = 1000$ is sufficient for the convergence in the regime of interest here.
For the momentum cutoff, as the same as that in the pure three-body interaction case, $\Lambda$ is also taken as $10k_{\rm F}$ in the ensuing calculations.

\begin{figure}
  \includegraphics[width=0.45\textwidth]{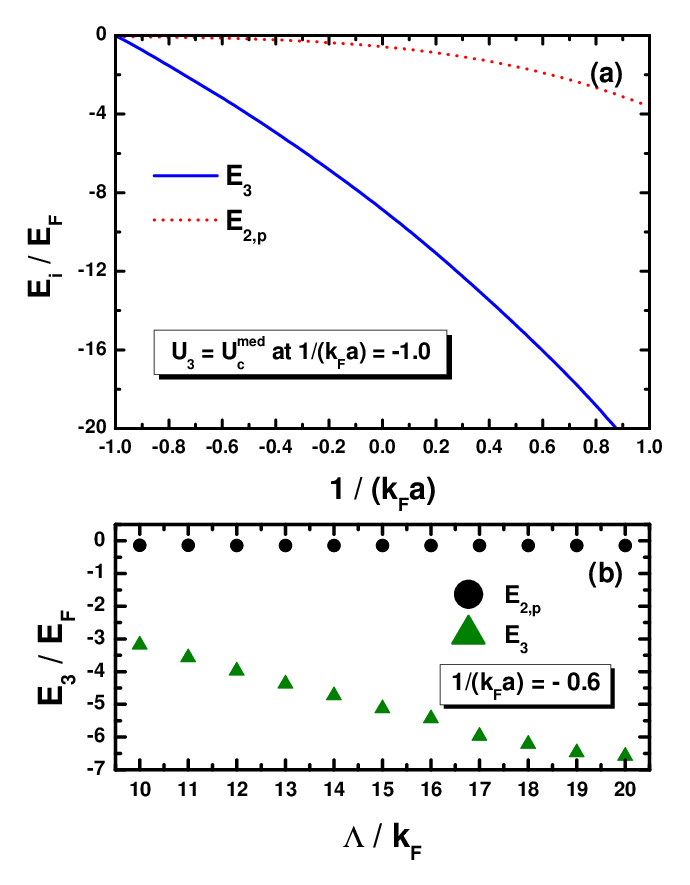}
  \caption{(a) In-medium three-body energy $E_3$ as a function of $1/(k_{\rm F}a)$ solved from Eq.~\eqref{3beq}.
  The three-body coupling constant ${U}_3$ is taken as in-medium critical coupling $U_c^{\rm med}=0.986 U_{\rm c}$ at $1/(k_Fa)=-1.0$.
 (b) Momentum cutoff dependence of $E_{2,p}$ and $E_3$ at $1/(k_{\rm F}a)=-0.6$. 
  }\label{fig:3}
\end{figure}

Figure \ref{fig:3} shows the numerical solution of $E_3$ as a function of $1(k_{\rm F}a)$,
where $U_3$ is taken as
the in-medium critical coupling $U_c^{\rm med}$ for the formation of the in-medium three-body bound state $1/(k_{\rm F}a)=-1.0$.
From the result shown in Fig.~\ref{fig:3}, $E_3$ monotonically decreases due to the increasing two-body coupling strength.
In addition, as shown in the lower panel of Fig.~\ref{fig:3}, where $1/(k_{\rm F}a)=-0.6$ is adopted, while $E_3$ exhibits a cutoff dependence, the qualitative behavior of the binding energies (e.g., $E_3<E_{2,p}$) is unchanged.
As a result, we conclude that $\Lambda/k_{\rm F}=10$ is sufficient for understanding physical properties of in-medium bound states in the present system.

\begin{figure}
  \includegraphics[width=0.45\textwidth]{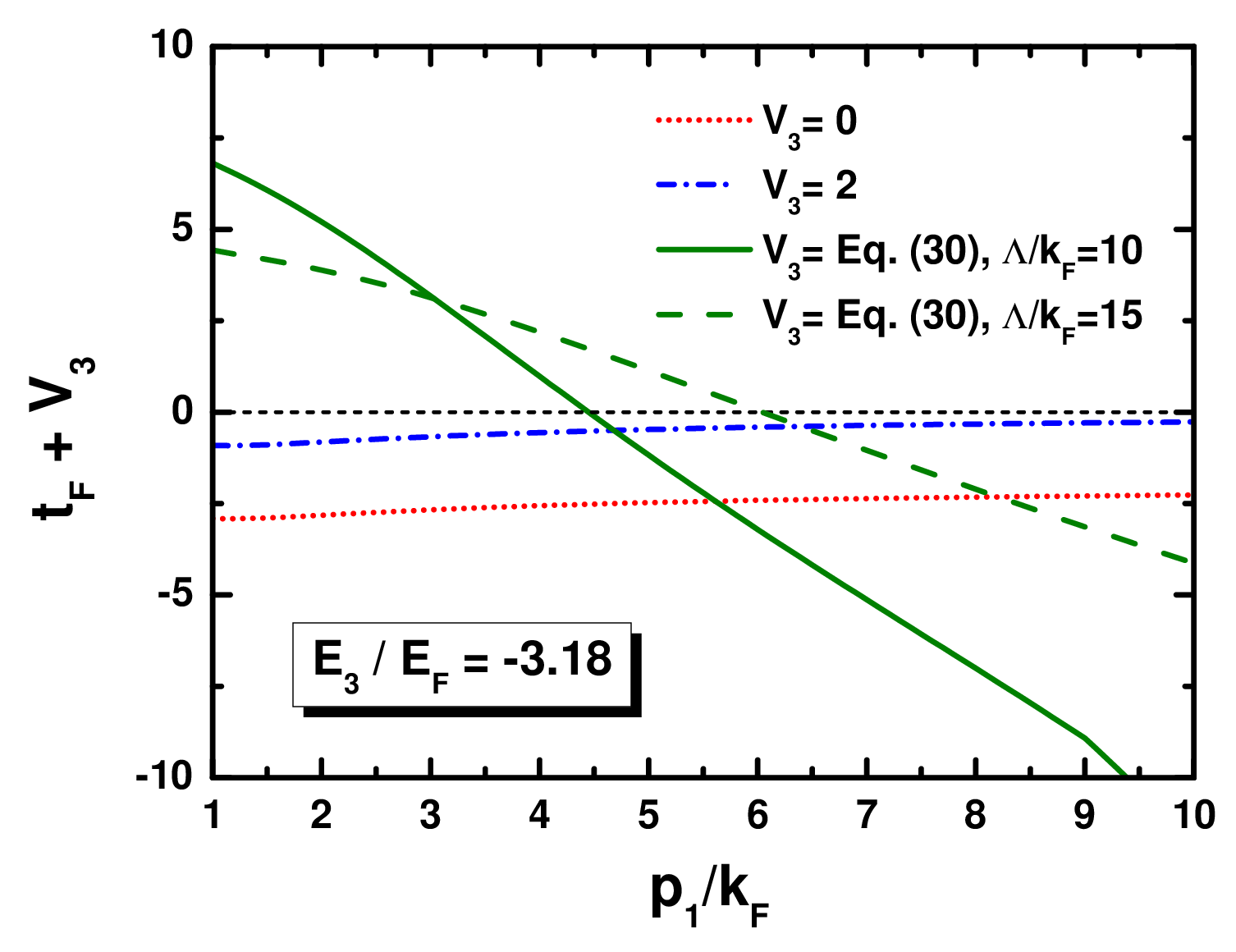}
  \caption{Dimensionless three-body interaction kernel $t_{\rm F}(p_1/k_{\rm F},p_2/k_{\rm F},E_3/E_{\rm F})+\mathcal{V}_3(p_1/k_{\rm F},p_2/k_{\rm F},E_3/E_{\rm F})$ as functions of $p_1$ with $p_2/k_{\rm F}=1$ and $E_3/E_{\rm F}=-3.18$ for different $\mathcal{V}_3$.
  The three-body coupling constant ${U}_3$ is taken as in-medium critical coupling $U_c^{\rm med}$ at $1/(k_Fa)=-1.0$.
  }\label{fig:2}
\end{figure}

To see the effect of the renormalized three-body coupling, $t_{\rm F}(p_1,p_2,E_3)+\mathcal{V}_3(p_1,p_2,E_3)$ defined in Eq.~\eqref{eq:v3} is shown as a function of $p_1$ for the cases with $\mathcal{V}_3(p_1,p_2,E_3)=0$, $\mathcal{V}_3(p_1,p_2,E_3)=2$ (i.e., the constant three-body coupling proposed in Ref.~\cite{Sekino2021Phys.Rev.A103.043307}), and the form of Eq.~\eqref{eqv3}, respectively, in Fig.~\ref{fig:2}, where we take $p_2=k_{\rm F}$ and $E_3=-3.18E_{\rm F}$.
In the case with $\mathcal{V}_3=2$, one can find the fermion-dimer repulsion induced by the three-body kernel $t_{\rm F}(p_1,p_2,E_3)$ is canceled by the constant coupling $\mathcal{V}_3=2$, when comparing it with the result of $\mathcal{V}_3=0$.
On the other hand, the momentum-dependent coupling $\mathcal{V}_3(p_1,k_{\rm F},E_3)$ generated by $U_3$ in Eq.~\eqref{eqv3} exhibits the low-energy attraction at $p_1\lesssim 5k_{\rm F}$, leading to the in-medium three-body bound state.
While it shows a cutoff dependence at large momenta due to the present form factor in the three-body interaction,
such a high-momentum repulsion is weakened when the cutoff increases.
This result is consistent with the cutoff dependence of $E_3$ shown in Fig.~\ref{fig:2}(b).
Eventually, even though the fermion-dimer repulsion is present in the high-momentum regime, the low-momentum attraction near the Fermi surface (i.e., $p_{1}\simeq k_{\rm F}$ and $p_{2}\simeq k_{\rm F}$) associated with $U_3$ induces the in-medium three-body bound states.

\begin{figure}
  \includegraphics[width=0.45\textwidth]{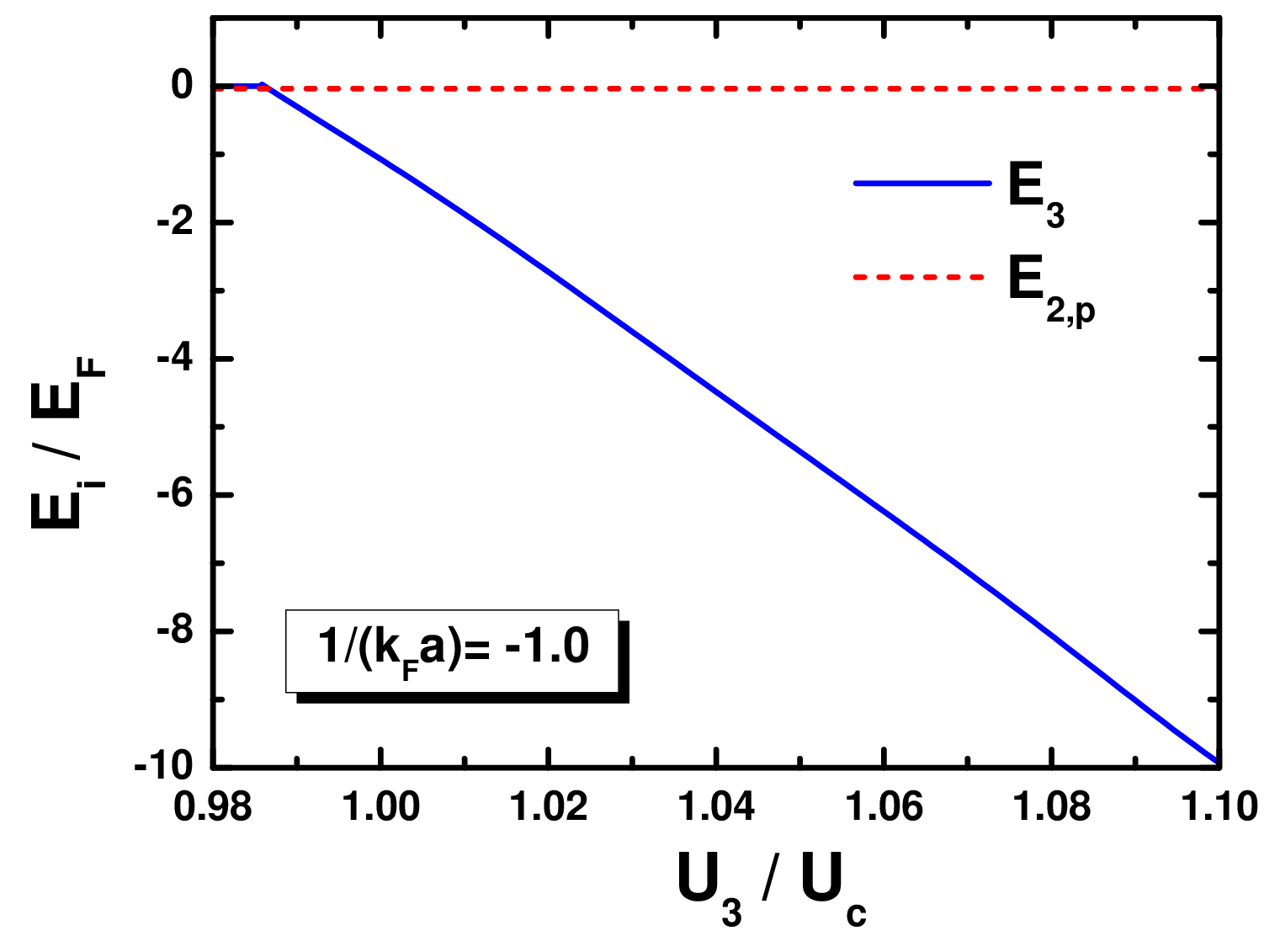}
  \caption{In-medium three-body energy $E_3$ as a function of dimensionless coupling constant of three-body interaction ${U}_3$ at $1/(k_{\rm F}a)=-1.0$ solved from Eq.~\eqref{3beq}.
  ${U}_3$ is normalized in the reference scheme of the critical coupling strength $U_c$.
  }\label{fig:4}
\end{figure}

To see the role of $U_3$ in detail,
$E_3$ as a function of three-body coupling strength ${U}_3$ at $1/(k_{\rm F}a)=-1.0$ solved from Eq.~\eqref{3beq} is also shown in Fig.~\ref{fig:4}.
${U}_3$ is normalized in the reference scheme of the critical coupling strength $U_c$.
From Fig.~\ref{fig:4}, it is seen that the three-body clusters gradually become more tightly bound with the increase of three-body coupling constant.
While pure three-body force case always exhibits squeezed Cooper triples, it is not always true in the presence of two-body attraction.
It is seen that $E_3$ becomes finite here even when $U_3/U_c<1$.
On the other hand, indeed, the definition of the boundary of the crossover between Cooper triple phase and squeezed one also involves an ambiguity.

\begin{figure}
  \includegraphics[width=0.45\textwidth]{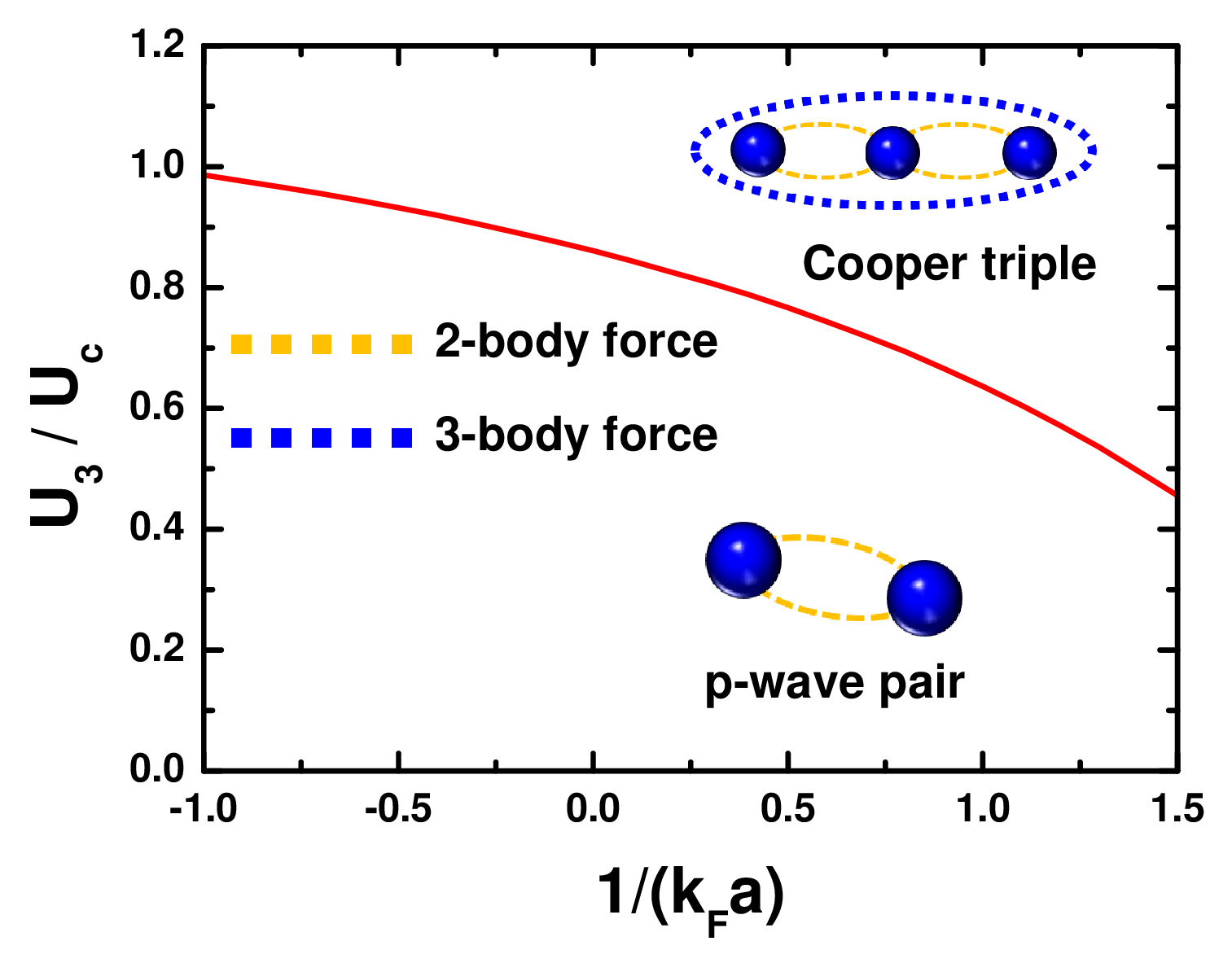}
  \caption{ 
  Phase diagram of $p$-wave pair phase ($|E_{2,p}|>|E_3|$) and Cooper triple phase ($|E_3|>|E_{2,p}|$) in the plane of ${U}_3/U_c$ and $1/(k_{\rm F}a)$.
  Namely, the Cooper triple ($p$-wave pair) phase is more stable in the regime above (below) the solid line.
  }\label{fig:5}
\end{figure}

For comparison, we also show $p$-wave Cooper pairing energies $E_{2,p}$ in Figs.~\ref{fig:3} and \ref{fig:4}, which can be obtained from the in-medium two-body equations for the $p$-wave pairing~\cite{Guo2022Phys.Rev.A106.043310}
\begin{align}
\label{eq:e2p}
    1+U_2\sum_{k}'\frac{k^2}{\xi_{k}+\xi_{-k}-E_{2,p}}=0.
\end{align}
In Fig.~\ref{fig:5},
we summarize the ground-state phase diagram of $p$-wave Cooper pairing states and the Cooper tripling state in the present model.
The phase boundaries are determined in such a way that 
the boundary between tripling and $p$-wave pairing is given by
$E_3=E_{2,p}$.
Such a phase diagram captures the competition between $p$-wave pairings and tripling.
Figure~\ref{fig:5} shows that the transition line between $p$-wave pair and Cooper triple phase monotonically decreases with the increasing two-body interaction.
However, this critical three-body coupling cannot trivially become zero at the strong-coupling limit [$1/(k_{\rm F}a)\rightarrow+\infty$], as we have figured out that the solution of three-body bound state cannot be found in the absence of three-body interaction~\cite{Guo2022Phys.Rev.A106.043310}.
Consequently, there is a threshold for the transition line at a certain two-body coupling strength [around $1/(k_{\rm F}a)\simeq 2$ in our calculation] due to the competition between the fermion-dimer repulsion and $U_3$ in the strong-coupling regime.
We note that such a threshold is not expected to be universal and would be associated with a detailed high-momentum structure of the interactions.
The quantitative investigation of such a behavior point and the properties in stronger coupling regimes are out of the scope in this paper.

While the values of $\Lambda$ and $U_3$ should be clarified to compare our results with the experiments,
our phase diagram would be useful to understand the qualitative features of systems with the coexistence of two- and three-body interactions.
Since the three-body ground-state energy depends on both the two- and three-body coupling constants, one can explore the competition between pairing and tripling phases by tuning the interactions in cold atomic systems.

\section{Summary and Perspectives}\label{sec:IV}

In this paper, we have investigated the in-medium three-body correlations (i.e., in the presence of Fermi sea) in one-dimensional spinless fermions with two- and three-body interactions.
We solve the in-medium three-body equation derived from the variational approach based on the generalized Cooper problem. 
In contrast to the previous works~\cite{Sekino2021Phys.Rev.A103.043307,Guo2022Phys.Rev.A106.043310}, where a specific form of the constant three-body interaction was phenomenologically introduced,
we have employed the antisymmetrized three-body interaction which involves a minimal momentum dependence as the leading-order contribution at low energy.

We first studied the simplified case with $U_2=0$, namely, the pure three-body interaction case.
The three-body energies have been obtained for both the in-medium and in-vacuum cases.
Unlike three-dimensional three-component Fermi gases, it has been found that the in-medium three-body bound state assisted by the Fermi-surface effect does not exist in the absence of the in-vacuum counterpart and the two-body interaction.
However, there is still a non-trivial regime corresponding to the squeezed Cooper-triple phase and moreover, the one which originates from the Fermi surface effect found in Ref.~\cite{Tajima2021Phys.Rev.A104.053328} in the presence of two-body attractions.
Similar to the BCS-BEC crossover in the three-dimensional $s$-wave superfluid Fermi gas~\cite{doi:10.1146/annurev-conmatphys-031113-133829,Strinati2018Phys.Rep.738.1--76,Ohashi2020Prog.Part.Nucl.Phys.111.103739}, the three-body cluster also undergoes a crossover from the Cooper tripling regime in the weak-coupling side to a regime of tightly bound trimers in the strong coupling side when the attractive interactions increase.

We have also investigated the general case with the coexistence of two- and three-body interactions.
In our previous work~\cite{Guo2022Phys.Rev.A106.043310}, the one-dimensional fermions were found to be stable against the three-body clustering when only two-body attractive $p$-wave interaction was considered.
Meanwhile, by including the dimensionless constant three-body coupling, it was found that an in-medium three-body state similar to a squeezed Cooper triple appears.
Similarly, in the present work, with further consideration of the antisymmetrized attractive three-body interaction, the stable three-body clusters survive as expected.
The in-medium three-body cluster is found to be more tightly bound with the increase of the three-body coupling strength.
Finally, we have featured a phase diagram consisting of the $p$-wave Cooper-pair and Cooper-triple phase in the plane of $p$-wave two-body coupling and three-body coupling strengths.
One can explore the competition between pairing and tripling phases by tuning the interactions in cold-atomic systems.

Our results would be useful for further investigation of unconventional superconductors and superfluids. 
Moreover, an in-medium three-body bound state with the existence of a non-negligible three-body interaction also paves a promising way for the study of higher-order clusters associated with the Fermi-surface effect.
The medium effect on bound trimers in higher dimensions such as the super Efimov state~\cite{Nishida2013PhysRevLett.110.235301} would also be an interesting topic.
More detailed studies on quantum correlations would also be worth investigating~\cite{PhysRevLett.130.253401}.

\begin{acknowledgments}

The authors are grateful to Kazuaki Takasan and RIKEN iTHEMS Non-Equilibrium Working group (NEW) for fruitful discussions.
Y.G. was supported by the RIKEN Junior Research Associate Program.
H.T. acknowledges the JSPS Grants-in-Aid for Scientific Research under Grants No.~18H05406, No.~22K13981, and No.~22H01158.

\end{acknowledgments}

\appendix
\begin{widetext}
\section{Expectation value of Hamiltonian}\label{appendixA}

In this appendix, we show the detailed expressions for the expectation values of each term in the Hamiltonian.
By applying the trial wave function~\eqref{twf} to the Hamiltonian~\eqref{hamiltonian}, the expectation values for the kinetic and two-body interaction parts are then obtained as~\cite{Guo2022Phys.Rev.A106.043310}
\begin{align}
    \langle \Psi_{3}\left| K\right|\Psi_{3} \rangle
    =\,&\sum_{p_1,p_2,p_3,p_1',p_2'.p_3'}'\Omega_{p_1,p_2}^*\Omega_{p_1',p_2'}(\xi_{p_1}+\xi_{p_2}+\xi_{p_3})\epsilon_{p_1,p_2,p_3}\epsilon_{p_1',p_2',p_3'}\delta_{p'_3,-p'_1-p'_2}\delta_{p_3,-p_1-p_2}\cr
    =\,&2\sum_{p_1,p_2}'(\xi_{p_1}+\xi_{p_2}+\xi_{-p_1-p_2})\Omega_{p_1,p_2}^*\left[\Omega_{p_1,p_2}+\Omega_{p_2,-p_1-p_2}+\Omega_{-p_1-p_2,p_1}\right],
\end{align}
with the Levi-Civita symbol $\epsilon_{p_1,p_2,p_3}$
and
\begin{align}
    \langle\Psi_{3}\left| V_2\right|\Psi_{3}\rangle =\,&\frac{U_2}{2}
    \sum_{k_1,k_2}'
    \sum_{k_1',k_2'}'
    \sum_{p_1,p_2,p_3}'
    \sum_{p_1',p_2',p_3'}'\left(\frac{k_1-k_2}{2}\right)\left(\frac{k_1'-k_2'}{2}\right)\Omega_{p_1,p_2}^*\Omega_{p_1',p_2'} \delta_{p_1+p_2,-p_3}\delta_{p_1'+p_2',-p_3'}\cr
    &\times\langle{\rm FS}|F_{p_1,p_2,p_3}B_{k_1,k_2}^\dag B_{k_1',k_2'}F_{p_1',p_2',p_3'}^\dag
    |{\rm FS}\rangle\nonumber\\
    \equiv\,&2v_{21}+v_{22},
\end{align}
respectively, where we defined
\begin{subequations}
\begin{align}
v_{21}=\frac{U_2}{2}\sum_{p_1,p_2,q}'\Omega_{p_1,p_2}^* 
\left[(p_1-p_2)(2q-p_1-p_2)\Omega_{-p_1-p_2,q}
+(2p_2+p_1)(2q+p_1)\Omega_{p_1,q}
+(-2p_1-p_2)(2q+p_2)\Omega_{p_2,q}
\right],
\end{align}
and
\begin{align}
v_{22}=\frac{U_2}{2}\sum_{p_1,p_2,q}'\Omega_{p_1,p_2}^* &\left[(p_1-p_2)(2q-p_1-p_2)\Omega_{q,-q+p_1+p_2}
+(2p_2+p_1)(2q+p_1)\Omega_{q,-q-p_1}\right.&\cr
&\left.
+(-2p_1-p_3)(2p_1+p_2)\Omega_{q,-q-p_2}
\right].
\end{align}
\end{subequations}
At last, the expectation value for the three-body interaction part reads
\begin{align}
&\langle\Psi_3\left|V_3\right|\Psi_3\rangle\nonumber\\
=\,&
U_3
\sum_{k_1,k_2,k_3}'
\sum_{k'_1,k'_2,k'_3}'
\sum_{p_1,p_2,p_3}'
\sum_{p'_1,p'_2,p'_3}'
\left(\frac{k_1-k_2}{2}\right)
\left(\frac{k_1'-k_2'}{2}\right)
\left(\frac{k_2-k_3}{2}\right)
\left(\frac{k_2'-k_3'}{2}\right)
\left(\frac{k_3-k_1}{2}\right)
\left(\frac{k_3'-k_1'}{2}\right)\nonumber\\
&\times
\Omega^\ast_{p_1,p_2}
\Omega_{p'_1,p'_2}
\delta_{k_1+k_2+k_3,k'_1+k'_2+k'_3} 
\delta_{p_1+p_2,-p_3}
\delta_{p'_1+p'_2,-p'_3}
\langle{\rm FS}\left|
c_{p_3} 
c_{p_2} 
c_{p_1} 
c_{k_1}^\dag 
c_{k_2}^\dag 
c_{k_3}^\dag 
c_{k'_3} 
c_{k'_2} 
c_{k'_1}  
c_{p'_1}^\dag 
c_{p'_2}^\dag 
c_{p'_3}^\dag 
\right|{\rm FS}\rangle\nonumber\\
=\,&\frac{9U_3}{16}
\sum_{p_1,p_2}'
\sum_{p'_1,p'_2}'
\left({p_1-p_2}\right)
\left({p_1'-p_2'}\right)
\left({p_1+2p_2}\right)
\left({p_1'+2p_2'}\right)
\left({2p_1+p_2}\right)
\left({2p_1'+p_2'}\right)
\Omega^\ast_{p_1,p_2}
\Omega_{p'_1,p'_2}.
\end{align}

\section{Derivation of Eq.~\eqref{closedeq}}\label{appendixB}

In this appendix, we show the detailed derivations of Eq.~\eqref{closedeq}.
The amplitude $\Omega_{p_1,p_2}$ can be expressed in terms of $\mathcal{A}(p_1,p_2)$, $\mathcal{B}(p_2)$, and $\mathcal{C}$  as,
\begin{align}
\label{eq:omega1}
    \Omega_{p_1,p_2}+\Omega_{p_2,p_3}+\Omega_{p_3,p_1}
    =-\frac{\frac{U_2}{2}\left[(p_1-p_2)\mathcal{B}(p_3)+(p_2-p_3)\mathcal{B}(p_1)+(p_3-p_1)\mathcal{B}(p_2)\right]
    }{2(\xi_{p_1}+\xi_{p_2}+\xi_{p_3}-E_3)}
+\frac{9U_3}{16}\mathcal{C}\frac{
\mathcal{X}_{p_1,p_2,p_3}}{2(\xi_{p_1}+\xi_{p_2}+\xi_{p_3}-E_3)}.
\end{align}
The above equation can be also recast into
\begin{align}\label{eqbp}
    \mathcal{B}(p_2)
    \left[1+\frac{U_2}{2}\sum'_{p_1}\frac{(p_3-p_1)^2}{2(\xi_{p_1}+\xi_{p_2}+\xi_{p_3}-E_3)}\right]
    =-U_2\sum'_{p_1}\frac{(p_2-p_3)(p_3-p_1)\mathcal{B}(p_1)
    }{2(\xi_{p_1}+\xi_{p_2}+\xi_{p_3}-E_3)}
+\frac{9U_3}{16}\mathcal{C}\sum'_{p_1}\frac{
\mathcal{X}_{p_1,p_2,p_3}
\left({p_3-p_1}\right)}{2(\xi_{p_1}+\xi_{p_2}+\xi_{p_3}-E_3)}.
\end{align}
By multiplying $\mathcal{X}_{p_1,p_2,p_3}\delta_{p_1+p_2+p_3,0}$  on both sides in Eq.~(\ref{eq:omega1}) and taking the momentum summation with respect to $p_1,p_2,p_3$, one has
\begin{align}
    -3\mathcal{C}=-\frac{U_2}{2}\sum'_{p_1,p_2}
    \frac{\mathcal{X}_{p_1,p_2,p_3}[(p_1-p_2)\mathcal{B}(p_3)
    +(p_2-p_3)\mathcal{B}(p_1)
    +(p_3-p_1)\mathcal{B}(p_2)]
    }{2(\xi_{p_1}+\xi_{p_2}+\xi_{p_3}-E_3)}
    +\frac{9U_3}{16}\mathcal{C}\sum'_{p_1,p_2}
    \frac{\mathcal{X}_{p_1,p_2,p_3}^2}{2(\xi_{p_1}+\xi_{p_2}+\xi_{p_3}-E_3)}.
\end{align}
Since the antisymmetric tensor 
$\mathcal{X}_{p_1,p_2,p_3}$
satisfies
\begin{align}
    \mathcal{X}_{p_2,p_1,p_3}=(p_2-p_1)(p_1-p_3)(p_3-p_2)=-(p_1-p_2)(p_2-p_3)(p_3-p_1)\equiv-\mathcal{X}_{p_1,p_2,p_3},
\end{align}
the first term in the right-hand side of the above equation can be recast into,
\begin{align}
&\sum'_{p_1,p_2}\frac{\mathcal{X}_{p_1,p_2,p_3}(p_2-p_3)\mathcal{B}(p_1)}{2(\xi_{p_1}+\xi_{p_2}+\xi_{p_3}-E_3)}
+\sum'_{p_1,p_2}\frac{\mathcal{X}_{p_1,p_2,p_3}(p_3-p_1)\mathcal{B}(p_2)}{2(\xi_{p_1}+\xi_{p_2}+\xi_{p_3}-E_3)}
+\sum'_{p_1,p_2}\frac{\mathcal{X}_{p_1,p_2,p_3}(p_1-p_2)\mathcal{B}(p_3)}{2(\xi_{p_1}+\xi_{p_2}+\xi_{p_3}-E_3)}
\cr
&\ \ =
3\sum'_{p_1,p_2}\frac{\mathcal{X}_{p_1,p_2,p_3}(p_2-p_3)\mathcal{B}(p_1)}{2(\xi_{p_1}+\xi_{p_2}+\xi_{p_3}-E_3)}.
\end{align}
In this way, we obtain
    \begin{align}\label{eqbp2}
    \mathcal{C}&=\left[1+\frac{3U_3}{16}\sum'_{p_1,p_2}
    \frac{\mathcal{X}_{p_1,p_2,p_3}^2}{2(\xi_{p_1}+\xi_{p_2}+\xi_{p_3}-E)}\right]^{-1}\frac{U_2}{2}\sum'_{p_1,p_2}
    \frac{(p_2-p_3)\mathcal{X}_{p_1,p_2,p_3}\mathcal{B}(p_1)
    }{2(\xi_{p_1}+\xi_{p_2}+\xi_{p_3}-E_3)}.
\end{align}
By combining Eqs.~\eqref{eqbp} and \eqref{eqbp2}, we obtain the closed equation for $\mathcal{B}(p)$ and $E_3$ as
\begin{align}
    &\mathcal{B}(p_2)
    \left[\frac{1}{U_2}+\sum'_{p_1}\frac{(p_1+p_2/2)^2}{\xi_{p_1}+\xi_{p_2}+\xi_{p_3}-E_3}\right]
    =\sum'_{p_1}\frac{(p_1+2p_2)(p_1+p_2/2)\mathcal{B}(p_1)
    }{\xi_{p_1}+\xi_{p_2}+\xi_{p_3}-E_3}\cr
&+\frac{\frac{9U_3}{16}\sum'_{p_1}\frac{
\mathcal{X}_{p_1,p_2,p_3}
\left({p_3-p_1}\right)}{2(\xi_{p_1}+\xi_{p_2}+\xi_{p_3}-E_3)}}{1+\frac{3U_3}{16}\sum'_{p_1,p_2}
    \frac{\mathcal{X}_{p_1,p_2,p_3}^2
    }{2(\xi_{p_1}+\xi_{p_2}+\xi_{p_3}-E_3)}}
    \sum'_{p_1,p_2}\frac{(p_2+p_1/2){\mathcal{X}_{p_1,p_2,p_3}}\mathcal{B}(p_1)}{2(\xi_{p_1}+\xi_{p_2}+\xi_{p_3}-E_3)}.
\end{align}
\end{widetext}


%

\end{CJK}
\end{document}